\documentclass[lettersize,journal]{IEEEtran}
\usepackage[utf8]{inputenc}
\usepackage[T1]{fontenc}
\usepackage{booktabs}
\usepackage{hyperref}
\usepackage{amsthm}
\usepackage{amsmath}
\usepackage[inline]{enumitem}
\usepackage{dirtytalk}
\usepackage{xspace}
\usepackage{amssymb}
\usepackage{graphicx}
\usepackage{upgreek}
\usepackage{fancyhdr}
\usepackage{tikz}
\usetikzlibrary{shapes, shapes.geometric, calc, arrows.meta, fit, positioning}
\usepackage{siunitx}
\usepackage[numbers]{natbib}

\usepackage[markup=underlined]{changes}
\definechangesauthor[name={JK}, color=purple]{JK}
\definechangesauthor[name={YS}, color=teal]{YS}
\definechangesauthor[name={WA}, color=blue]{WA}

\usepackage{todonotes}
\presetkeys{todonotes}{inline,fancyline}{}
\newcommand{\isofusa}{ISO~26262}
\newcommand{\isosotif}{ISO/PAS~21448}

\newcommand{\tab}{Table}

\newcommand{\fig}{Fig.}

\newcommand{\sect}{Section\xspace}
\newcommand{\sense}{\textsc{sense}\xspace}
\newcommand{\dc}{\textsc{plan}\xspace}
\newcommand{\act}{\textsc{act}\xspace}
\newcommand{\mdl}{\ensuremath{\mathcal{M}}}
\newcommand{\prf}{\ensuremath{\mathcal{P}}\xspace}
\newcommand{\require}{\ensuremath{\mathcal{R}}}
\newcommand{\predcoll}{\ensuremath{\mathit{collision}}\xspace}
\newcommand{\dl}{\text{\upshape\textsf{d{\kern-0.03em}L}}\xspace}
\newcommand{\odd}{\ensuremath{\mathit{ODD}}\xspace}
\newcommand{\sen}{\ensuremath{\mathit{sense}}\xspace}
\newcommand{\ctrl}{\ensuremath{\mathit{plan}}\xspace}
\newcommand{\plant}{\ensuremath{\mathit{act}}\xspace}
\newcommand{\xp}{\ensuremath{x_p}\xspace}
\newcommand{\vp}{\ensuremath{v_p}\xspace}
\newcommand{\hatxp}{\ensuremath{\hat{x}_p}\xspace}
\newcommand{\hatvp}{\ensuremath{\hat{v}_p}\xspace}
\newcommand{\areq}{\ensuremath{a^{req}}\xspace}
\newcommand{\amin}{\ensuremath{a^{min}}\xspace}
\newcommand{\amax}{\ensuremath{a^{max}}\xspace}
\newcommand{\range}{\ensuremath{\mathit{range}}\xspace}
\newcommand{\safe}{\ensuremath{\mathit{safe}}\xspace}
\newcommand{\gsnparskip}{\\[0.35\baselineskip]}
\newcommand{\goalprobability}{G4\xspace}
\newcommand{\goalformal}{G5\xspace}
\newcommand{\goalsense}{G6\xspace}
\newcommand{\goalctrl}{G7\xspace}
\newcommand{\goalact}{G8\xspace}
\newcommand{\stratexposure}{S2\xspace}
\newcommand{\stratreqbreak}{S3\xspace}
\newcommand{\stratreqfull}{S4\xspace}
\newcommand{\pctl}[1]{\mathbb{P}_{#1}}

\tikzstyle{GSNgoal} = [rectangle, minimum width=3cm, minimum height=1cm, align=left, draw=black]
\tikzstyle{GSNstrategy} = [trapezium, trapezium left angle=70, trapezium right angle=110, minimum width=3cm, minimum height=1cm, align=left, draw=black]
\tikzstyle{GSNsolution} = [circle, minimum width=1cm, minimum height=1cm, align=left, draw=black]
\tikzstyle{GSNcontext} = [rounded rectangle, minimum width=3cm, minimum height=1cm, align=left, draw=black]
\tikzstyle{GSNassumption} = [ellipse, minimum width=3cm, minimum height=1cm, align=left, draw=black, label={[anchor=north west, below right]south east:A}]
\tikzstyle{GSNjustification} = [ellipse, minimum width=3cm, minimum height=1cm, align=left, draw=black, label={[anchor=north west, below right]south east:J}]
\tikzstyle{GSNundeveloped} = [diamond, minimum width=.5cm, minimum height=.5cm, draw=black, anchor=north]

\tikzstyle{GSNsupportedBy} = [-{Triangle[fill=black, width=3mm, length=5mm]}]
\tikzstyle{GSNcontextOf} = [-{Triangle[open, width=3mm, length=5mm]}]

\newtheorem{requirement}{\require\ignorespaces}
\newtheorem{model}{\mdl\ignorespaces}

\title{A Formal-Methods Approach to Provide Evidence in Automated-Driving Safety Cases}
\author{Jonas~Krook*, Yuvaraj~Selvaraj*, Wolfgang~Ahrendt and Martin~Fabian\thanks{* Equal contribution.}\thanks{This work was supported by the Wallenberg AI, Autonomous Systems and Software Program (WASP) funded by the Knut and Alice Wallenberg Foundation and by FFI, VINNOVA under grant 2017-05519, \textit{Automatically Assessing Correctness of Autonomous Vehicles -- Auto-CAV}.}\thanks{Jonas Krook and Yuvaraj Selvaraj are with Zenseact, 417 56 Gothenburg, Sweden,
and also with the Department of Electrical Engineering, Chalmers
University of Technology, 412 96 Gothenburg, Sweden (e-mail:
\{firstname.lastname\}@zenseact.com).}\thanks{Wolfgang Ahrendt is with the Department of Computer Science and Engineering, Chalmers University of Technology, 412 96 Gothenburg, Sweden.}
\thanks{Martin Fabian is with the Department of Electrical Engineering, Chalmers University of Technology, 412 96 Gothenburg, Sweden.}}
\date{}

\begin{document}

\maketitle
\thispagestyle{fancy}
\pagestyle{empty}
\fancyhf{}
\fancyfoot[L]{\footnotesize This work has been submitted to the IEEE for possible publication.\\Copyright may be transferred without notice, after which this version may no longer be accessible.}
\renewcommand{\headrulewidth}{0pt}
\renewcommand{\footrulewidth}{0pt}

\begin{abstract}
    The safety of automated driving systems must be justified by convincing arguments and supported by compelling evidence to persuade certification agencies, regulatory entities, and the general public to allow the systems on public roads. This persuasion is typically facilitated by compiling the arguments and the compelling evidence into a \emph{safety case}. Reviews and testing, two common approaches to ensure correctness of automotive systems cannot explore the typically infinite set of possible behaviours. In contrast, \emph{formal methods} are exhaustive methods that can provide mathematical proofs of correctness of models, and they can be used to prove that formalizations of functional safety requirements are fulfilled by formal models of system components. This paper shows how formal methods can provide evidence for the correct break-down of the functional safety requirements onto the components that are part of feedback loops, and how this evidence fits into the argument of the safety case. If a proof is obtained, the formal models are used as requirements on the components. This structure of the safety argumentation can be used to alleviate the need for reviews and tests to ensure that the break-down is correct, thereby saving effort both in data collection and verification time.    
\end{abstract}


\section{Introduction}
\IEEEPARstart{A}{utomated Driving Systems}~(ADS) relieve the human driver from the driving task and let the driver engage in other activities while travelling~\cite{SAE:J3016:21}. Among several potential benefits of ADS, one in particular is to prevent accidents caused by driver errors and thereby increase traffic safety~\cite{EU:19b:short}. An indication that such potential traffic safety benefit exists is provided by extrapolation from previous experience with driver assistance systems~\cite{EugBraFraRotSolRob:13}, and by analyses showing that ADS have the potential to prevent or reduce the severity of several accident scenarios involving human drivers~\cite{LubJepRanFreBarOst:18}. However, these studies only apply to situations that human drivers do not handle safely, and \emph{accidents} involving human drivers; they do not provide insights into situations that human drivers already handle safely.

ADS are designed to take over operation of the \emph{Dynamic Driving Task}~(DDT) in environments that are included in a specific \emph{Operational Design Domain}~(ODD)~\cite{SAE:J3016:21}. To provide a net increase in traffic safety, an ADS must perform the DDT such that it overall is safer than a human driver in all the environments of the ODD. It must also limit operation to environments included in the ODD, or risk unsafe behavior.
As human drivers in general have a very low failure rate~\cite{KalPad:16} and as the ADS cannot expect human supervision or intervention in its ODD, the ADS becomes highly safety critical.

Two pertinent problems arise when developing these safety-critical ADS: \begin{enumerate*}[label=(\roman*)]
    \item the development methods and processes that are applied must ensure safety of ADS~\cite{KooWag:17}, and
    \item this fact must be supported by compelling evidence to persuade certification agencies, regulatory entities, and the general public of the safety of ADS. In essence, safety of the ADS must be demonstrated, rather than assumed until proven unsafe~\cite{KooKanBla:19}.
\end{enumerate*}

ADS have a bidirectional interaction with their surrounding environment; that is, an ADS must adapt its behavior to the environment, and the environment behavior is affected by the ADS's actions. This paper considers ADS that can be represented by an architecture with three components as shown in \fig~\ref{fig:ads:arch}. The \sense component senses and perceives the environment, the \dc component is responsible for decisions on when and how to act, and the \act component executes the decisions using the respective actuators. These components interact with the environment, here represented by the ODD, in a feedback loop.

\tikzset{
     block/.style={rectangle, draw, text width=4.5em,
                   text centered, rounded corners, minimum height=2.3em},
     arrow/.style={-{Stealth[]},thick}
     }

\begin{figure}[!tb]
\centering
\begin{tikzpicture}
    \node [block] (sense) {\textsc{\sense}};
    \node [block,right=0.5cm of sense] (dc)  {\textsc{\dc}};
    \node [block,right=0.5 cm of dc] (act)  {\textsc{\act}};
    \node [block, below=0.4 cm of dc] (odd1) {\textsc{odd}};
    \node [draw, thick, dashed, fit={(sense) (dc) (act)}, inner xsep=0.85 cm, inner ysep=0.22 cm, label={ADS}] (temp) {};
    
    \draw [arrow] (sense) -- (dc);
    \draw[arrow] (dc) -- (act);
    \draw [arrow] (act.east) -| ($(act.east)!.5!(temp.east)$) |- (odd1.east);
    \draw [arrow] (odd1.west) -| ($(sense.west)!.5!(temp.west)$) |- (sense.west);
\end{tikzpicture}
\caption{A simplified architecture for an ADS.}
\label{fig:ads:arch}
\end{figure}
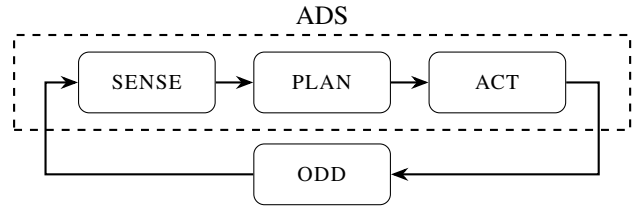

The feedback loop between an ADS and its environment means that compelling safety evidence must be gathered in closed-loop conditions, and the compiled evidence must be shown to be representative of all the environments in the ODD. One way to accomplish both of these points is to perform real-world driving that covers the entire ODD, and drive enough distance such that safety can be evaluated. However, testing in real-world driving conditions as the only means for aiding safety-critical development and producing compelling evidence for safety is infeasible for all but trivial ODDs~\cite{KalPad:16,KooKanBla:19,KooWag:16}. 
To overcome this obstacle, the process of evidence collection often follows a divide-and-conquer approach. This is done by first breaking down the system-level ADS requirements to its components, and then further to smaller elements until the effort to support each requirement with compelling evidence is acceptable~\cite{ISO:21448:22, ISO:26262:18}. A benefit of this approach is that the evidence can be collected by methods that are specialized for the specific type of requirement, but which may not be feasible for system-level requirements. A drawback, however, is that fulfillment of the component-level requirements must now imply the fulfillment of the system-level requirements in the entire ODD. To ensure that there are no gaps in the broken-down requirements, verification must still be performed for the complete ADS, but to a lesser extent.

%

\emph{Reviews} and \emph{Testing} are two often recommended methods to ensure correctness of electronic automotive features~\cite{ISO:26262:18}. Both of these methods are used to find faults or insufficiencies 
throughout
the development process. They are complementary and often find different kinds of issues. However, neither is exhaustive, mainly because reviews are laborious, and because testing cannot explore the typically infinite set of possible behaviors in its entirety. These challenges are exacerbated for feedback systems, especially when discrete decisions are taken, and when errors take long time to propagate into failures. For instance, to avoid collisions with pedestrians, an ADS must approach road-side pedestrians with a suitable speed so that it can guarantee to stop safely, should the pedestrian step out into the road. A bad decision by the ADS can cause a collision several seconds or minutes later; deciding to pass the pedestrian although it is too close to the road will erroneously remove the braking option, but this error will not become apparent until, and if, the pedestrian enters the road.


\emph{Formal methods} are a category of methods that can \emph{prove} and ensure correctness of feedback-system models with respect to the requirements. In contrast to testing, these methods are exhaustive and provide evidence that no faults or insufficiencies are present in the component model, at least not w.r.t.\ the specified behaviour. They are also automatic or semi-automatic, so they typically require less labor than reviews. 
This paper shows how formal methods can be used to justify that components relying on feedback fulfill their requirements, and thereby give compelling evidence for the safety of the component. Furthermore, this paper demonstrates how formal methods can be used to close the gap between the ADS system requirements and the broken-down component requirements. This is done by showing how the formalization of the system-level safety requirements of the ADS puts assumptions and verification conditions on the different components, and on the ODD. The assumptions thus obtained also give a formal description of the conditions that must be fulfilled for the formal proof to be valid, and this paper puts forth an argument that those conditions in certain instances may be evaluated in open-loop settings, thereby considerably decreasing the verification effort. Both the above contributions are illustrated and argued based on a relevant example from the industry.

\section{Related Work}

\isofusa\ lists formal methods as techniques for ensuring dependability on the software architectural and unit design level~\cite{ISO:26262:18}.
At these levels, the software is typically considered as open-loop input/output systems, and this is also the intended setting in the standard. In this context, formal methods can provide evidence that the software is dependable~\cite{DenPaiPoh:12}. Formal methods are not considered at other levels of the design in the standard, and hence there are no recommendations regarding formal methods applied to feedback systems.

Formal methods have been used successfully in the automotive domain to prove that complex feedback systems are correct with respect to safety in a given environment~\cite{LooPlaNis:11, NilOzaTopMur:12, WonTopMur:13, KorDolVanRenHee:18, SelAhrFab:22, KroSveLiFenFab:19}. However, these works do not demonstrate how the artifacts from the formal methods contribute to a convincing argument that safety is achieved. 

Previous research has established that there are opportunities for using evidence from formal methods to convincingly argue for safety in all levels of the design of safety-critical systems~\cite{GalIwuMcDToy:08, Rus:10}. Moreover, it is argued that the assumptions on the environment are an important part of the model, and their inclusion allow more focus on the evidence and validation instead of ensuring that the break-down of requirements is correct~\cite{GalIwuMcDToy:08, Rus:10}; an argument which is supported by the contributions in \sect~\ref{sec:discussion}. There is also a challenge in how to treat probabilistic requirements when employing formal methods~\cite{Rus:10}, which is also addressed in \sect~\ref{sec:fm:pcs}.

When using formal methods to argue that a system is safe, it is imperative that both the underlying formalism and the tools that are being used are correct. Obviously, there must be a convincing argument that this is indeed the case, but such argument is out of scope of this paper as it is typically available from literature associated with the respective methods. There are many aspects to consider with respect to the correctness of the formalism and any tool being used, and a generic argument that captures all these aspects is provided by \citeauthor*{HabKel:09}~\cite{HabKel:09}.

\section{Safety Case}

A safety-critical system, such as an ADS, must behave such that safe operation is ensured in its entire ODD, where, commonly, \say{safe} is taken to mean that there is \emph{an absence of an unreasonable risk of harm}. When a safety-critical ADS is developed, it must be ensured that its behavior indeed is safe, but it must also be justified by compelling evidence that the risk of harm is low enough. This latter part is required to persuade certification agencies, regulatory entities, and the general public.

The justification of an ADS's safety can be compiled into a \emph{safety case}, which is \say{a structured argument, supported by a body of evidence that provides a compelling, comprehensible and valid case that a system is safe for a given application in a given operating environment}~\cite{MoD:00-56:17}. There are three principal elements in a safety case: requirements, arguments, and evidence~\cite{KelWea:04}. The safety case approach has been used in many safety critical industries to demonstrate safety, and is also recommended by automotive safety standards such as the \isofusa~\cite{ISO:26262:18} and \isosotif~\cite{ISO:21448:22}.

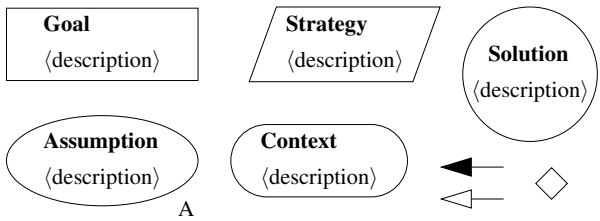
\begin{figure}[t]
\centering
\resizebox{0.9\linewidth}{!}{
\begin{tikzpicture}
    \node [GSNsolution, inner sep=0.025cm] (solution) {\textbf{\ \ Solution}\gsnparskip $\langle$description$\rangle$};
    \coordinate (cnw) at (solution.north -| solution.west);
    \node [GSNstrategy, left=1.25cm of cnw, anchor=north east] (strategy) {\textbf{Strategy }\gsnparskip $\langle$description$\rangle$};
    \node [GSNgoal, left=of strategy] (goal) {\textbf{Goal}\gsnparskip $\langle$description$\rangle$};
    \node [GSNassumption, below=1.25cm of goal.south west, anchor=west, inner sep=0.05cm] (assumption) {\textbf{Assumption}\gsnparskip $\langle$description$\rangle$};
    \node [GSNcontext, right=0.5cm of assumption] (context) {\textbf{Context}\gsnparskip $\langle$description$\rangle$};
    \coordinate [right=0.5cm of context] (temp);
    \coordinate [below=0.1cm of temp] (i1);
    \coordinate [right=of i1] (i2);
    \coordinate [below=0.5cm of i1] (i3);
    \coordinate [right=of i3] (i4);
    \draw[GSNsupportedBy](i2) -- (i1);
    \draw[GSNcontextOf](i4) -- (i3);
    \node [GSNundeveloped, right=0.75cm of i2, anchor=north] (undeveloped) {};
\end{tikzpicture}}
\caption{The core elements of a GSN goal structure. The line with a solid arrowhead denotes a 
\emph{SupportedBy}
relationship and the line with the hollow arrowhead denotes an 
\emph{InContextOf}
relationship. The diamond indicates an undeveloped element.}
\label{fig:GSN:core}
\end{figure}

Since a safety case is used to demonstrate that a product is safe, it is imperative that its structure is clear, comprehensible, and accurate. The \emph{Goal Structuring Notation}~(GSN) is a standardized graphical argument notation~\cite{KelWea:04, GSN:21}, which can be used to structure a safety case. It explicitly documents the individual elements of a safety argument and their relationships to the gathered evidence. GSN defines core elements, two types of relationships between the core elements, and an undeveloped element decorator, as shown in \fig~\ref{fig:GSN:core}. The two relationships \emph{SupportedBy} and \emph{InContextOf} declare a relationship between a source element and a target element. The elements are linked together in a logical structure known as the \emph{GSN goal structure}, which is a directed acyclic graph. The top-level goal in a goal structure is gradually refined through a series of more detailed goals until a direct link to evidence is made~\cite{KelWea:04}. The undeveloped element decorator indicates that a line of argument has not been developed in the current context.

\section{Formal Methods}
\label{sec:fm}

Formal methods are a class of mathematically rigorous techniques and tools used to specify, design, verify, and synthesize components, mainly by mechanizing rigorous reasoning about correctness of these components~\cite{NASA:16}. As such, the field is very broad, and a wide plethora of tools are available for many different parts of the development process and at different levels of abstraction. This paper is concerned with formal methods applied to the function layer of the ADS, i.e., details of the hardware and the software are abstracted away.

Formal methods are based on languages with formal syntax and semantics that leave no room for ambiguity. Common to all classes of formal methods is that a requirement on the system, in this case an ADS, is formalized into a \emph{specification} that details correct behavior of the system. In the case of input/output systems, the formal specification relates the required output to certain input, but in this paper the specification details the allowed behavior over time for feedback systems. 

Since the safety requirements for an ADS describe disallowed and mandated behaviors over time, logical formalisms that support modelling and reasoning about properties with respect to time, such as \emph{Linear Temporal Logic}~(LTL)~\cite{BaiKat:08} and \emph{differential dynamic logic}~(\dl)~\cite{Pla:18}, are typically used to formalize the requirements. Often safety requirements are characterized as \say{nothing bad shall happen}~\cite{BaiKat:08}, which is easily formalized using the modal operators to describe necessity in LTL and in \dl. For instance, $\Box \phi$ in LTL asserts that the property $\phi$ \emph{always} holds and $[\mdl]\phi$ in \dl asserts that after \emph{all} behaviours of model $\mdl$, the property $\phi$ holds.

Given such a formal specification, \emph{formal verification} and \emph{formal synthesis} provide evidence of correctness of formal models of feedback systems in the form of a machine checked formal proof of the fulfillment of the specification. For formal verification, all parts in \fig~\ref{fig:ads:arch}, and their possible interactions, are modelled in a formal language, and then the formal verification tool attempts to prove that the formal model fulfills the formal specification. For formal synthesis, the goal is instead to automatically construct, typically, a model of \dc such that the resulting feedback system is guaranteed to fulfill the formal specification. In the case of synthesis, the models of \sense, \act, and the ODD are commonly referred to as the \emph{assumptions}, whereas the required behavior of the feedback system is referred to as the \emph{guarantees}. This distinction is not as common in the case of formal verification, but in this paper the modelled parts will be referred to as the assumptions of the system regardless of method.

\section{Proposed Safety Argument Approach}\label{sec:prop:app}

For feedback systems like ADS, the break-down of requirements onto components can be difficult since requirements on the system-level refer to behaviors of the closed-loop system in the context of the ODD, whereas the component requirements specify the behavior at the interfaces between the components. The fulfillment of the component requirements must imply the fulfillment of the system requirements; if this is not the case, there is either a behavior disallowed by the system requirements that is allowed by the component requirements, or there is a behavior mandated by the system requirements that is not mandated by the component requirements. When the involved requirements are safety critical, this potential discrepancy between the behaviors that are disallowed or mandated at different levels might lead to an unsafe ADS. Compelling evidence that this potential discrepancy does not decrease the safety to an unacceptable level must be dealt with in the safety case~\cite{ISO:26262:18, BerJohSodNilTry:15}.

The approach in this paper uses formal methods to prove that the break-down has no such discrepancy. The system requirement is formalized into a guarantee to be fulfilled, and the formal system model is composed of the formal models of the components and the ODD. The approach is to use these models as specifications for the components. The component specifications can be developed into formal contracts~\cite{SanDamPas:12} that provide unambiguous requirements for the separate components. By this decoupling, specialized methods can be used to develop and verify the broken-down requirements. Hence, if the guarantee is proven to be fulfilled by the formal models, and if the verification of the components indicate that they fulfill their formal specifications, then the system requirement is fulfilled in the ODD, and there is compelling evidence for this fact in the form of a formal proof.

For instance, the guarantee could be a formalization of \say{there are no impacts with pedestrians}. Likely, for the ADS to fulfill this guarantee, the component \act must be able to provide some deceleration below some minimal level. More specifically, assume that the component \dc requests an acceleration $\areq$ in a certain range, then \act must ensure that the true acceleration $a$ of the vehicle fulfills $a \leq \areq$. Then the property $a \leq \areq$ is considered the specification for the \act component.


In the case that the system requirement is quantitative, specifying a probability or an occurrence rate of an event, then the formal guarantee is specified such that the event must not occur. In the break-down process, the models of the components are assigned a probability or occurrence rate with which they may be violated, such that the cumulative violations of the assumptions in the formal model does not exceed what is allowed by the system requirement.

The approach is illustrated in \sect~\ref{sec:fm:pcs} with an example where a safety requirement with respect to pedestrians is broken down to safety requirements on the closed-loop system. It is then shown how the artifacts of the formalization of these requirements can be used to structure an argument in a safety case. The argument is illustrated graphically by a GSN goal structure in \fig~\ref{fig:gsn:argument} to make the relations between requirements and arguments clear.

\section{Precautionary Safety and Risk Norm}\label{sec:pcs:qrn}

The example in this paper is based on the concept of \emph{Precautionary Safety}~(PCS)~\cite{RodKiaBra:21}. PCS attempts to ensure safety by adjusting an ADS's behavior based on its capabilities, external conditions, and expected exposure to incidents. The notion of safety in this context follows the concept of \emph{Quantitative Risk Norm}~(QRN)~\cite{WarSkoThoJohBraGyl:20}, where certain accident types with certain severity (in terms of human injury) are assigned a minimum allowed mean time between accidents (risk norm). Based on an assessment of Swedish national accident injury data for pedestrians and an estimation of total hours driven, an acceptable mean time between accidents
can be established. The severity is highly correlated with impact speed~\cite{PedScuSleMohHydJar:04}, so the allocated risk norms are in \tab~\ref{tab:risk:norm} given based on impact speed.

\begin{table}[tb]
    \centering
    \caption{Risk Norms for Pedestrians. Different Mean Time Between Failures for Different Accident Severities.}
    \label{tab:risk:norm}
    \begin{tabular}{@{\,}cr@{\,}}
        \begin{tabular}{c}Impact speed\\{}[\si{km/h}]\end{tabular} & Risk norm [\si{h}] \\
        \midrule
        $<$ 10 & \num{100 000} \\
        $[10, 20)$ & \num{1 000 000} \\
        $[20, 30)$ & \num{10 000 000} \\
        $[30, 40)$ & \num{100 000 000} \\
        $\geq$ 40 & \num{1 000 000 000} \\
    \end{tabular}
\end{table}

The ADS is intended to operate up to a speed of \SI{70}{km/h} on urban roads, and up to \SI{100}{km/h} on highways, so these are the environments that make up the ODD. The allowed failure rate of the ADS depends both on the risk norm and on the incident exposure rate in the ODD, i.e., the mean time between the occurrence of pedestrians. This exposure rate is different for different road segments, as pedestrians are more likely to appear on the road in low-speed urban settings than on highways with free-flowing traffic. The assumed exposure rates for urban and highway driving are given in \tab~\ref{tab:exposure} (c.f. \citeauthor{RodKiaBra:21}~\cite{RodKiaBra:21}).

\begin{table}[tb]
    \centering
    \caption{Exposure Levels with Respect to Speed and Road Type}
    \label{tab:exposure}
    \begin{tabular}{@{\,}crr@{\,}}
         Speed & \multicolumn{2}{c}{Exposure [\si{h}]} \\
         \cmidrule(lr{.16667em}){2-3}
         [\si{km/h}] & Urban Roads & Highways \\
         \midrule
         \num{30} & \num{100} & \num{100 000} \\
         \num{50} & \num{1 000} & \num{1 000 000} \\
         \num{70} & \num{10 000} & \num{10 000 000} \\
         \num{100} & - & \num{10 000 000} \\
    \end{tabular}
\end{table}



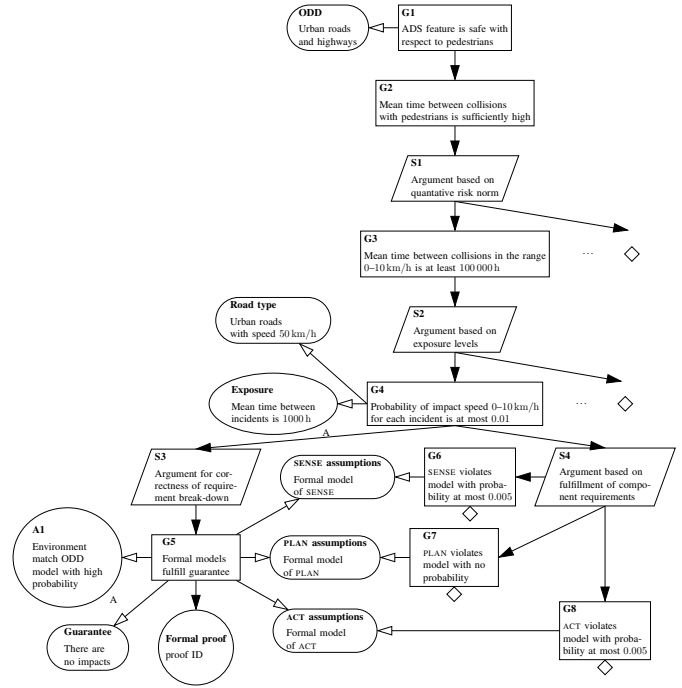
\begin{figure}[tb]
    \centering
    \resizebox{\linewidth}{!}{\begin{tikzpicture}
        \node [GSNgoal] (g1) {\textbf{G1}\gsnparskip ADS feature is safe with\\respect to pedestrians};
        \node [GSNcontext, left=of g1] (c11) {\textbf{ODD}\gsnparskip Urban roads\\and highways};
        \node [GSNgoal, below=of g1] (g2) {\textbf{G2}\gsnparskip Mean time between collisions\\with pedestrians is sufficiently high};
        \node [GSNstrategy, below=of g2] (s1) {\textbf{S1}\gsnparskip Argument based on\\quantative risk norm};
        \node [GSNgoal, below=of s1] (g3) {\textbf{G3}\gsnparskip Mean time between collisions in the range\\\SIrange{0}{10}{km/h} is at least \SI{100000}{h}};
        \node [right=of g3] (dots) {\ldots};
        \node [right=of dots] (temp1) {};
        \node (temp) at (g3.north -| temp1) {};
        \node [GSNstrategy, below=of g3] (s3) {\textbf{\stratexposure}\gsnparskip Argument based on\\exposure levels};
        
        \node [GSNgoal, below=of s3] (g8) {\textbf{\goalprobability}\gsnparskip Probability of impact speed \SIrange{0}{10}{km/h}\\for each incident is at most \num{0.01}};
        \node [right=of g8] (dots8) {\ldots};
        \node [right=of dots8] (temp81) {};
        \node (temp8) at (g8.north -| temp81) {};
        \node [GSNassumption, left=of g8] (a81) {\textbf{Exposure}\gsnparskip Mean time between\\incidents is \SI{1000}{h}};
        \node [GSNcontext, above=of a81] (c81) {\textbf{Road type}\gsnparskip Urban roads\\with speed \SI{50}{km/h}};
        \node [GSNcontext, below left=of g8, anchor=north] (c7) {\textbf{\sense assumptions}\gsnparskip Formal model\\of \sense};
        \node [GSNstrategy, left=of c7] (s2) {\textbf{\stratreqbreak}\gsnparskip Argument for cor-\\rectness of require-\\ment break-down};
        \node [GSNgoal, right=of c7] (g7) {\textbf{\goalsense}\gsnparskip \sense violates\\model with proba-\\bility at most \num{0.005}};
        \node [GSNstrategy, right=of g7] (s4) {\textbf{\stratreqfull}\gsnparskip Argument based on\\fulfillment of compo-\\nent requirements};
        \node [GSNgoal, below=of s2] (g9) {\textbf{\goalformal}\gsnparskip Formal models\\fulfill guarantee};
        \node [GSNassumption, left=of g9] (a91) {\textbf{A1}\gsnparskip Environment\\match ODD\\model with high\\probability};
        \node [GSNcontext, right=of g9] (c5) {\textbf{\dc assumptions}\gsnparskip Formal model\\of \dc};
        \node [GSNgoal, right=of c5] (g5) {\textbf{\goalctrl}\gsnparskip \dc violates\\model with no\\probability};
        \node [GSNcontext, below=of c5] (c6) {\textbf{\act assumptions}\gsnparskip Formal model\\of \act};
        \node (g6) at (c6 -| s4) [GSNgoal] {\textbf{\goalact}\gsnparskip \act violates\\model with proba-\\bility at most \num{0.005}};
        \node [GSNsolution, below=of g9] (sol) {\textbf{Formal proof}\gsnparskip proof ID};
        \node [GSNcontext, left=of sol] (guarantee) {\textbf{Guarantee}\gsnparskip There are\\no impacts};
    
        \node [GSNundeveloped] (g6u) at (g6.south) {};
        \node [GSNundeveloped] (g5u) at (g5.south) {};
        \node [GSNundeveloped] (g7u) at (g7.south) {};
        \node [GSNundeveloped, anchor=center] (s1u) at (temp |- g3) {};
        \node [GSNundeveloped, anchor=center] (s3u) at (temp8 |- g8) {};
    
        \draw [GSNsupportedBy] (g1.south) -- (g2.north);
        \draw [GSNsupportedBy] (g2.south) -- (s1.north);
        \draw [GSNsupportedBy] (s1.south) -- (g3.north);
        \draw [GSNsupportedBy] (g3.south) -- (s3.north);
        \draw [GSNsupportedBy] (s3.south) -- (g8.north);
        \draw [GSNsupportedBy] (g8.south) -- (s2.north);
        \draw [GSNsupportedBy] (g8.south) -- (s4.north);
        \draw [GSNsupportedBy] (s2.south) -- (g9.north);
        \draw [GSNsupportedBy] (g9.south) -- (sol);
        \draw [GSNsupportedBy] (s4.south) -- (g5.east);
        \draw [GSNsupportedBy] (s4.west) -- (g7.east);
        \draw [GSNsupportedBy] (s4.south) -- (g6.north);
    
        \draw [GSNsupportedBy] (s1.south) -- (temp);
        \draw [GSNsupportedBy] (s3.south) -- (temp8);
    
        \draw [GSNcontextOf] (g1) -- (c11);
        \draw [GSNcontextOf] (g8) -- (a81);
        \draw [GSNcontextOf] (g8.west) -- (c81);
        \draw [GSNcontextOf] (g9) -- (a91);
        \draw [GSNcontextOf] (g9) -- (c5);
        \draw [GSNcontextOf] (g9) -- (c7);
        \draw [GSNcontextOf] (g9) -- (c6);
        \draw [GSNcontextOf] (g9) -- (guarantee);
        \draw [GSNcontextOf] (g5) -- (c5);
        \draw [GSNcontextOf] (g7) -- (c7);
        \draw [GSNcontextOf] (g6) -- (c6);
    \end{tikzpicture}}
    \caption{GSN illustrating the argument made in this paper. A bigger version is available at \url{https://doi.org/10.5281/zenodo.7142341}.}
    \label{fig:gsn:argument}
\end{figure}

The QRN argument that the ADS feature is sufficiently safe with respect to pedestrians is illustrated graphically by the goals G1, G2, and G3 in \fig~\ref{fig:gsn:argument}. G1 represents the top-level safety requirement that the ADS feature shall be safe with respect to pedestrians. G1 is then made more specific in G2 by relating safety to its definition of having sufficiently high mean time between accidents. G2 is fulfilled via the strategy S1, which argues that the ADS is sufficiently safe whenever the risk norms in \tab~\ref{tab:risk:norm} are fulfilled, as exemplified by G3. Note that the four goals corresponding to the other risk norms in parallel to G3 are not shown. G3 is broken down based on exposure level, corresponding to the different road types. Each combination of exposure level and risk norm gives rise to an impact probability per event, where the maximum allowed impact probability in \goalprobability is calculated by the ratio of exposure and risk norm. Not shown in parallel to \goalprobability in the branch rooted in G3 are the six other exposure classes from \tab~\ref{tab:exposure}. All other goals in parallel to G3 are supported by an analogous argument structure. This paper and the PCS paper
~\cite{RodKiaBra:21}
agrees on the safety case so far, but differs below \goalprobability.

In \citeauthor{RodKiaBra:21}~\cite{RodKiaBra:21}, an already implemented reactive collision-avoidance module's capability to avoid collisions by braking is determined in simulations. The capability is presented as the probability of a certain impact speed given the speed before braking starts. The mean time between certain impact speeds for certain roads is then calculated as the ratio of the exposure on that road and the probability of that impact speed given the speed of the road. The mean time between impact speeds is then compared to the QRN to determine the maximum allowed speed that the ADS may drive on that road; if the mean time between accidents of a certain impact speed for a certain initial speed is lower than the QRN, then the ADS may not drive that fast on that road. For instance, exposure on urban roads with speed \SI{70}{km/h} is \SI{10 000}{h}, and the probability of an impact speed of \SIrange{10}{20}{km/h} for a speed of \SI{60}{km/h} is \SI{5}{\percent}. The ratio is \SI{200 000}{h}, considerably less than the QRN of \SI{1 000 000}{h}, so the ADS cannot be allowed to drive at \SI{60}{km/h} on urban roads with a speed of \SI{70}{km/h}. Thus, safety is ensured by taking precaution; the maximum speed is adapted based on the expected capability of mitigating impacts and the exposure to incidents.

However, this way of applying the PCS concept assumes an already implemented ADS. If it is desired to drive at \SI{70}{km/h} on urban roads, PCS indicates that the probability of impact speeds of \SIrange{10}{20}{km/h} ought to be at most \SI{1}{\percent}. One way of reaching such performance would be to iteratively implement improvements and simulate and test until the capability is sufficiently safe for driving at \SI{70}{km/h}. This paper shows how formal methods can be used to instead break down the requirement to components to achieve a systematic development process. It is also shown how the break-down through formal methods helps with structuring an argument supporting the fulfillment of the safety case.

The main idea of PCS still applies; the ADS ought to take precaution by considering what \emph{might} happen, not solely what \emph{is} happening, to avoid unacceptable risk of harm. Formal methods explore the entire state space of the formal model in order to prove the guarantee, so they fit well as a tool to ensure that all possible events are considered by the ADS.

\section{Formal Methods in the Safety Case}
\label{sec:fm:pcs}

To ease the development effort, the goal \goalprobability in \fig~\ref{fig:gsn:argument} may be broken down to \goalsense, \goalctrl, and \goalact that are assigned to the respective components in \fig~\ref{fig:ads:arch}, and in that way separate the concerns for design and verification of the different components. This is a typical way to combat complexity by allowing application of specialized tools and methods to the design and verification of the components~\cite{ISO:26262:18,SanDamPas:12}. As stated earlier, these components' goals must imply the goal \goalprobability in \fig~\ref{fig:gsn:argument} lest the risk of harm may be unacceptable.

Formal methods can provide proofs as evidence that, in the context of the ODD, the component models fulfill the guarantee. However, the guarantee is a qualitative logical formula, whereas \goalprobability is a quantitative goal with a probability. To deal with this, the strategy for fulfilling \goalprobability is split into one qualitative part and one quantitative part, as can be seen in \fig~\ref{fig:gsn:argument} where \goalprobability is supported by the two strategies \stratreqbreak and \stratreqfull. The idea here is to disregard the probability in \goalprobability and formalize the remainder as the guarantee, and then find a formal model that satisfies that guarantee. To reintroduce the probability, it is finally argued that the goal \goalprobability is fulfilled as long as the assumptions in the formal model are violated with at most the same probability as in \goalprobability.

The entire argument hinges on the correctness of the formal model with respect to the guarantee, and this is captured by \goalformal. It is made explicit with the context relation that the formal model is composed of the assumptions on the ODD and the models of the components \sense, \dc, and \act, and the guarantee which correctness is evaluated against. The sole evidence that \goalformal is fulfilled is provided by the machine checked formal proof.

Now assume that the behavior of the components \sense, \dc, and \act fulfill their corresponding formal assumptions with probabilities such that the entire formal model is fulfilled with a probability of at least \num{0.99}. Then \goalprobability is fulfilled because the guarantee that ``no impacts occur'' is violated at most with a probability of \num{0.01}, which in turn means that impacts in the range \SIrange{0}{10}{km/h} can occur at most with a probability of \num{0.01}.

Therefore, \goalprobability is broken down into \goalsense, \goalctrl, and \goalact, each detailing the probability of the assumed behaviors of \sense, \dc, and \act, respectively, are being violated. Here, the probabilities are assigned arbitrarily, but it is ensured that their sum
does not exceed the probability of \goalprobability. It can be argued via strategy \stratreqbreak that the break-down of \goalprobability to \goalsense, \goalctrl, and \goalact is correct, and it can be argued via strategy \stratreqfull that \goalprobability is fulfilled because \goalsense, \goalctrl, and \goalact are fulfilled. The component models can serve as formal specifications for the individual components, and relevant standards may be used to develop and verify the three components according to these specifications~\cite{ISO:21448:22, ISO:26262:18}.

To be a bit more specific, it is now illustrated with more detail what the context of \goalformal may look like. 
Since the intention is to show how a formal-methods approach could be used in the safety case and not discuss any specific formal method in detail, the approach in this paper is illustrated with abstract artifacts and simple models for the three components in \fig~\ref{fig:ads:arch}. The first type of artifact, the guarantee \require\ref{req:top:goal}, is the formal specification based on \goalprobability.

\begin{requirement}\label{req:top:goal}
There are no impacts with pedestrians. 
\end{requirement}

\require\ref{req:top:goal} can be formalized in different ways depending on the formalism used. For example, \require\ref{req:top:goal} can be formalized in LTL using the formula $\Box \lnot \predcoll$ where \predcoll is a predicate describing the undesirable property of the occurrence of an impact; and in \dl using the formula $[\mdl](\lnot \predcoll)$, where \mdl\ is the formal model. 

Irrespective of the formalism used, formal models of the components are required in order to either synthesize a controller that guarantees \require\ref{req:top:goal} or verify that a given design fulfills \require\ref{req:top:goal}. Thus, the second type of artifact is formal models of the ADS components in \fig~\ref{fig:ads:arch}. As for the formal specification, the ADS components can be modelled in different ways depending on the formalism of choice. 
To formalize the components, parameters about the ADS vehicle and pedestrians are considered, as shown in \tab~\ref{tab:sys:par}.   
\begin{table}[!tb]
    \centering
    \caption{Parameters Considered in the ADS Model}
    \label{tab:sys:par}
    \begin{tabular}{@{\,}cr@{\,}}
        \begin{tabular}{c}Parameter\end{tabular} & Description \\
        \midrule
         $x$, $v$, $a$ &  position, velocity, acceleration of ADS vehicle\\
         \amin, \amax &  minimum, maximum acceleration of ADS vehicle\\
         $\delta$ & actuator disturbance \\
         \xp, \vp & true position, lateral velocity of pedestrian \\
         \hatxp, \hatvp &  estimated position, lateral velocity of pedestrian\\
         $\epsilon$ & sensor disturbance \\
    \end{tabular}
\end{table}


This paper considers a \dc component with a safety controller such as the one in \citeauthor{SelAhrFab:22}~\cite{SelAhrFab:22} to guarantee safety. For the sake of brevity, \mdl\ref{mdl:ctrl} presents a very abstract model of \dc. The required acceleration \areq is set to the minimum acceleration \amin if the predicate $\lnot \safe$ is satisfied. This predicate can be used to describe decision-making conditions such as checking if there is some choice of $\areq \in [\amin, \amax]$ such that, later on, stopping before the pedestrian is infeasible. 

\begin{model}[\ctrl]\label{mdl:ctrl} $\quad\lnot \safe \to \areq = \amin$
\end{model}

Of course, to prove that \mdl\ref{mdl:ctrl} fulfills \require\ref{req:top:goal}, certain assumptions must be made on the other components. Typically such assumptions are identified as a result of the formal modelling and analysis. In this context, 
consider \mdl\ref{mdl:odd}, \mdl\ref{mdl:sense}, and \mdl\ref{mdl:plant} as the assumptions on the ODD, \sense, and \act, respectively. The assumption on the ODD describes that the ADS vehicle velocity $v$ is non-negative, and it defines the limits on pedestrian velocity $\vp$.

\begin{model}[\odd]\label{mdl:odd} $\quad v \ge 0 \land 0 \le \vp \le 10$
\end{model}

The range of allowed values for $\vp$ defines what may happen in the environment, and the exhaustiveness of formal methods make sure that all different combinations with all different timings are evaluated. A controller that fulfills the requirement in the presence of $\vp$ certainly takes precaution for what might happen, and not only reacts to what is happening.

\mdl\ref{mdl:sense} describes that, if the true position $\xp$ of the pedestrian is within the detection range of the sensor, then the error in the estimated position of the pedestrian $(\hatxp - \xp)$ is at most $\epsilon$. Furthermore, if $\xp$ is in front of the ADS vehicle, then $\hatxp$ is also estimated to be in front.

\begin{model}[\sen]\label{mdl:sense}
\begin{equation*}
    \left(\xp \le \range \to \hatxp - \xp \leq \epsilon\right)\, \land\, \left(\xp \ge x \to \hatxp \ge x\right) 
\end{equation*}
\end{model}
The tolerances make it possible to assign probabilities to the fulfillment of the specifications. The exhaustiveness of formal methods evaluates all combinations, so the exact probability distribution does not need to be known.

The assumptions on the \act described by \mdl\ref{mdl:plant} state that if the requested acceleration \areq is within the bounds, then the tracking error tolerance between actual acceleration and requested acceleration is at most $\delta$. 

\begin{model}[\plant]\label{mdl:plant} $\quad 
    \amin \le \areq \le \amax \to a \le \areq + \delta $
\end{model}

Assume that the formal model~\mdl\ composed by \mdl\ref{mdl:ctrl} -- \mdl\ref{mdl:plant} is correct with respect to the guarantee \require\ref{req:top:goal}, and that there exists a formal proof \prf that this is indeed the case. The proof \prf provides enough evidence that \goalformal in \fig~\ref{fig:gsn:argument} is fulfilled. 

For \goalctrl to be fulfilled, the realized controller in the component \dc must fulfill the behavior specified by \mdl\ref{mdl:ctrl}. This may be assured, for instance, by following the recommendations in \isofusa~\cite{ISO:26262:18} to achieve a fault tolerant realization. Since \mdl\ref{mdl:ctrl} is a simple condition relating inputs to outputs, much of its verification can be performed in open loop, which results in less effort to collect evidence that \goalctrl is fulfilled.

This paper does not consider strategies to validate the ODD, so in A1 in \fig~\ref{fig:gsn:argument}, \mdl\ref{mdl:odd} is considered an assumption in the argument, which means that there is no justification or evidence that \mdl\ref{mdl:odd} is fulfilled. Obviously, to ensure real-life safety, this assumption must be validated for the roads that the ADS vehicle is allowed to drive on. That endeavor may benefit greatly from having a formalized formulation of assumed properties of the ODD.

The model \mdl\ref{mdl:sense} specifies the required behavior of \sense for the guarantee to be fulfilled, and \goalsense details the probability with which this behavior may be violated. This requirement can, to a large extent, be verified in open loop on recorded data, provided that the behavior in the recordings adhere to the assumptions of \mdl\ref{mdl:odd}. This could provide a substantial benefit since the data may be collected before all components are realized, and because the recorded data is still relevant after implementation has changed in the components. This is not the case for the approach by \citeauthor{RodKiaBra:21}~\cite{RodKiaBra:21} when assessing the capability of the complete vehicle. Furthermore, the verification of \sense does not need to ensure that there are enough outcomes with different impact speeds, as the requirement is independent of the closed-loop outcome. This last point can potentially save effort both in data collection and verification time.

The last model, \mdl\ref{mdl:plant}, which specifies the acceleration tracking performance, must still be verified in closed-loop conditions. However, the break-down of requirement \goalprobability into \goalact makes the verification independent of the other components, which may simplify the verification method. To realize \sense and \act below the goals \goalsense and \goalact, the development process could, for instance, employ \isosotif\ and \isofusa.


\section{Discussion}
\label{sec:discussion}
The approach in \sect~\ref{sec:prop:app} is one possible argument structure to include formal proofs of correctness as evidence in safety cases. Whether the approach is beneficial depends on the overhead of the formal methods in comparison to the verification effort, and as such, the approach may be beneficial for some systems, and not for others, and it might be beneficial for only a subset of the requirements in one system.

The benefits are also dependent on the formal modelling and the system decomposition. Some system decompositions may not be possible to formally model in a given formalism, and the formal specifications of a component might fail to be verifiable, or exceedingly hard to verify. Such complications will trigger redesigns that could be costly.

On the other hand, a notable benefit is that the presented approach in itself is agnostic to the chosen methods and technologies with which the components are realized. Admittedly, some might be more amenable to be used in conjunction with the formal specifications.

Furthermore, if formal methods are employed at an early stage in the project, they may catch inconsistencies and design flaws that would be costly to find during system-level testing. The process of formalization of requirements can itself be beneficial, and the approach detailed in this paper allows more utilization of such work.

The proposed approach uses a strategy to split a quantitative goal into one qualitative and one quantitative part. The qualitative part disregards the probabilities involved and employs formal methods to provide evidence to show that the qualitative part is fulfilled. However, there exists formal approaches that can prove correctness of stochastic systems such as \emph{probabilistic model checking}~\cite{KwiNorPar:22} where quantitative extensions of temporal logic are used to specify quantitative properties. While it is beneficial to investigate the suitability of such approaches to provide evidence to quantitative goals in the safety argument, they are not considered in this paper.

\section{Conclusions}

This paper presents an approach to structure the safety case for employing formal methods in safety cases for automated driving systems. The artifacts from the formal analysis are used as formal specifications on the components of the system, thus providing a break-down of safety requirements to individual components. The formal proof provided by the formal method is used as evidence that the break-down is correct. If the safety requirement specify a quantitative target, then this is handled by assigning quantitative targets for the specifications on the components, in parallel to the formal model.

This approach gives several potential benefits by limiting the effort of safety assurance, and by breaking down requirements on a complex system into well defined and unambiguous specifications on individual components. First of all, less effort is needed to verify that the break-down of the requirement is correct, as this is proven by the formal method. Second, the formal specifications on the components provide separation of the concerns, and may also allow for verification in open-loop settings instead of closed-loop settings for some components, which leads to more flexibility and less effort to verify the individual components. Case studies of complex systems are needed to validate this approach.

\bibliographystyle{IEEEtranN}
\bibliography{krook-abrv, krook-common, krook-bibtex}

\begin{thebibliography}{32}
\providecommand{\natexlab}[1]{#1}
\providecommand{\url}[1]{#1}
\csname url@samestyle\endcsname
\providecommand{\newblock}{\relax}
\providecommand{\bibinfo}[2]{#2}
\providecommand{\BIBentrySTDinterwordspacing}{\spaceskip=0pt\relax}
\providecommand{\BIBentryALTinterwordstretchfactor}{4}
\providecommand{\BIBentryALTinterwordspacing}{\spaceskip=\fontdimen2\font plus
\BIBentryALTinterwordstretchfactor\fontdimen3\font minus
  \fontdimen4\font\relax}
\providecommand{\BIBforeignlanguage}[2]{{%
\expandafter\ifx\csname l@#1\endcsname\relax
\typeout{** WARNING: IEEEtranN.bst: No hyphenation pattern has been}%
\typeout{** loaded for the language `#1'. Using the pattern for}%
\typeout{** the default language instead.}%
\else
\language=\csname l@#1\endcsname
\fi
#2}}
\providecommand{\BIBdecl}{\relax}
\BIBdecl

\bibitem[{SAE J3016\char`_202104}(2021)]{SAE:J3016:21}
{SAE J3016\char`_202104}, ``Taxonomy and definitions for terms related to
  driving automation systems for on-road motor vehicles,'' {SAE} Int., Tech.
  Rep., Apr. 2021.

\bibitem[EU:(2019)]{EU:19b:short}
``Regulation {(EU)} 2019/2144 of the {European Parliament},'' \textsc{url:}
  \url{http://data.europa.eu/eli/reg/2019/2144/oj}, Dec. 2019.

\bibitem[Eugensson et~al.(2013)Eugensson, Br\"{a}nnstr\"{o}m, Frasher, Rothoff,
  Solyom, and Robertsson]{EugBraFraRotSolRob:13}
A.~Eugensson, M.~Br\"{a}nnstr\"{o}m, D.~Frasher, M.~Rothoff, S.~Solyom, and
  A.~Robertsson, ``Environmental, safety, legal and societal implications of
  autonomous driving systems,'' in \emph{Int. Tech. Conf. Enhanc. Saf. Veh.},
  vol. 334, 2013.

\bibitem[Lubbe et~al.(2018)Lubbe, Jeppsson, Ranjbar, Fredriksson, B\"{a}rgman,
  and \"{O}stling]{LubJepRanFreBarOst:18}
N.~Lubbe, H.~Jeppsson, A.~Ranjbar, J.~Fredriksson, J.~B\"{a}rgman, and
  M.~\"{O}stling, ``Predicted road traffic fatalities in germany: The potential
  and limitations of vehicle safety technologies from passive safety to highly
  automated driving,'' in \emph{Int. Res. Counc. Biomech. Inj.}, Sep. 2018, pp.
  17--52.

\bibitem[Kalra and Paddock(2016)]{KalPad:16}
N.~Kalra and S.~M. Paddock, \emph{Driving to Safety: How Many Miles of Driving
  Would It Take to Demonstrate Autonomous Vehicle Reliability?}\hskip 1em plus
  0.5em minus 0.4em\relax RAND Corporation, 2016.

\bibitem[Koopman and Wagner(2017)]{KooWag:17}
P.~Koopman and M.~Wagner, ``Autonomous vehicle safety: An interdisciplinary
  challenge,'' \emph{{IEEE} Intell. Transp. Syst. Mag.}, vol.~9, no.~1, pp.
  90--96, Jan. 2017.

\bibitem[Koopman et~al.(2019)Koopman, Kane, and Black]{KooKanBla:19}
P.~Koopman, A.~Kane, and J.~Black, ``Credible autonomy safety argumentation,''
  in \emph{27th Saf.-Crit. Syst. Symp.}, Feb. 2019.

\bibitem[Koopman and Wagner(2016)]{KooWag:16}
P.~Koopman and M.~Wagner, ``Challenges in autonomous vehicle testing and
  validation,'' \emph{{SAE} Int. J. Transp. Saf.}, vol.~4, no.~1, pp. 15--24,
  2016, {SAE} World Congr. Exhib.

\bibitem[{ISO/PAS 21448:2022}(2022)]{ISO:21448:22}
{ISO/PAS 21448:2022}, ``Road vehicles -- safety of the intended
  functionality,'' {ISO}, Tech. Rep., Jun. 2022.

\bibitem[{ISO 26262:2018}(2018)]{ISO:26262:18}
{ISO 26262:2018}, ``Road vehicles -- functional safety,'' {ISO}, Tech. Rep.,
  Dec. 2018.

\bibitem[Denney et~al.(2012)Denney, Pai, and Pohl]{DenPaiPoh:12}
E.~Denney, G.~Pai, and J.~Pohl, ``Heterogeneous aviation safety cases:
  Integrating the formal and the non-formal,'' in \emph{{IEEE} 17th Int. Conf.
  Eng. Complex Comput. Syst.}, 2012, pp. 199--208.

\bibitem[Loos et~al.(2011)Loos, Platzer, and Nistor]{LooPlaNis:11}
S.~M. Loos, A.~Platzer, and L.~Nistor, ``Adaptive cruise control: Hybrid,
  distributed, and now formally verified,'' in \emph{FM 2011: Formal Methods},
  ser. LNCS, 2011, pp. 42--56.

\bibitem[Nilsson et~al.(2012)Nilsson, \"{O}zay, Topcu, and
  Murray]{NilOzaTopMur:12}
P.~Nilsson, N.~\"{O}zay, U.~Topcu, and R.~M. Murray, ``Temporal logic control
  of switched affine systems with an application in fuel balancing,'' in
  \emph{Am. Control Conf.}, Jun. 2012, pp. 5302--5309.

\bibitem[Wongpiromsarn et~al.(2013)Wongpiromsarn, Topcu, and
  Murray]{WonTopMur:13}
T.~Wongpiromsarn, U.~Topcu, and R.~M. Murray, ``Synthesis of control protocols
  for autonomous systems,'' \emph{Unmanned Syst.}, vol.~01, no.~01, pp. 21--39,
  2013.

\bibitem[Korssen et~al.(2018)Korssen, Dolk, {Van De Mortel-Fronczak}, Reniers,
  and Heemels]{KorDolVanRenHee:18}
T.~Korssen, V.~Dolk, J.~{Van De Mortel-Fronczak}, M.~Reniers, and M.~Heemels,
  ``Systematic model-based design and implementation of supervisors for
  advanced driver assistance systems,'' \emph{{IEEE} Trans. Intell. Transp.
  Syst.}, vol.~19, no.~2, pp. 533--544, Feb. 2018.

\bibitem[Selvaraj et~al.(2022)Selvaraj, Ahrendt, and Fabian]{SelAhrFab:22}
Y.~Selvaraj, W.~Ahrendt, and M.~Fabian, ``Formal development of safe automated
  driving using differential dynamic logic,'' \emph{{IEEE} Trans. Intell.
  Veh.}, pp. 1--12, 2022.

\bibitem[Krook et~al.(2019)Krook, Svensson, Li, Feng, and
  Fabian]{KroSveLiFenFab:19}
J.~Krook, L.~Svensson, Y.~Li, L.~Feng, and M.~Fabian, ``Design and formal
  verification of a safe stop supervisor for an automated vehicle,'' in
  \emph{Int. Conf. Robot. Autom.}, May 2019, pp. 5607--5613.

\bibitem[Galloway et~al.(2008)Galloway, Iwu, McDermid, and
  Toyn]{GalIwuMcDToy:08}
A.~Galloway, F.~Iwu, J.~McDermid, and I.~Toyn, ``On the formal development of
  safety-critical software,'' in \emph{First {IFIP} Conf. Verif. Softw.:
  Theories, Tools, Exp. 2005}, ser. LNCS, vol. 4171, 2008, pp. 362--373.

\bibitem[Rushby(2010)]{Rus:10}
J.~Rushby, ``Formalism in safety cases,'' in \emph{Mak. Syst. Safer}, ser.
  SCSC, 2010, pp. 3--17, saf.-Crit. Syst. Symp.

\bibitem[Habli and Kelly(2009)]{HabKel:09}
I.~Habli and T.~Kelly, ``A generic goal-based certification argument for the
  justification of formal analysis,'' \emph{Electron. Notes Theor. Comput.
  Sci.}, vol. 238, no.~4, pp. 27--39, Sep. 2009, first Workshop Certif.
  Saf.-Crit. Softw. Control. Syst. (SafeCert 2008).

\bibitem[{Ministry of Defence}(2017)]{MoD:00-56:17}
{Ministry of Defence}, ``Defence standard 00-56 part 1: Safety management
  requirements for defence systems,'' Tech. Rep., Feb. 2017.

\bibitem[Kelly and Weaver(2004)]{KelWea:04}
T.~Kelly and R.~Weaver, ``The goal structuring notation -- a safety argument
  notation,'' in \emph{Int. Conf. Depend. Syst. Netw.}, Jul. 2004, workshop
  Assur. Cases.

\bibitem[ACWG(2021)]{GSN:21}
ACWG, ``Goal structuring notation community standard,'' \textsc{url:}
  \url{https://scsc.uk/r141C:1}, SCSC, Tech. Rep., May 2021.

\bibitem[{NASA}(2016)]{NASA:16}
{NASA}, ``What is formal methods?'' \textsc{url:}
  \url{https://shemesh.larc.nasa.gov/fm/fm-what.html} (accessed 2022-07-26),
  Apr. 2016.

\bibitem[Baier and Katoen(2008)]{BaiKat:08}
C.~Baier and J.-P. Katoen, \emph{Principles of Model Checking}.\hskip 1em plus
  0.5em minus 0.4em\relax {MIT} Press, 2008.

\bibitem[Platzer(2018)]{Pla:18}
A.~Platzer, \emph{Logical Foundations of Cyber-Physical Systems}.\hskip 1em
  plus 0.5em minus 0.4em\relax Springer, 2018.

\bibitem[Bergenhem et~al.(2015)Bergenhem, Johansson, S\"{o}derberg, Nilsson,
  Tryggvesson, T\"{o}rngren, and Ursing]{BerJohSodNilTry:15}
C.~Bergenhem, R.~Johansson, A.~S\"{o}derberg, J.~Nilsson, J.~Tryggvesson,
  M.~T\"{o}rngren, and S.~Ursing, ``How to reach complete safety requirement
  refinement for autonomous vehicles,'' in \emph{Crit. Automot. Appl.: Robust.
  Saf.}, Sep. 2015.

\bibitem[Sangiovanni-Vincentelli et~al.(2012)Sangiovanni-Vincentelli, Damm, and
  Passerone]{SanDamPas:12}
A.~Sangiovanni-Vincentelli, W.~Damm, and R.~Passerone, ``Taming {Dr.}
  {Frankenstein:} contract-based design for cyber-physical systems,''
  \emph{Eur. J. Control}, vol.~18, no.~3, pp. 217--238, 2012.

\bibitem[Rodrigues~de Campos et~al.(2021)Rodrigues~de Campos, Kianfar, and
  Br\"{a}nnstr\"{o}m]{RodKiaBra:21}
G.~Rodrigues~de Campos, R.~Kianfar, and M.~Br\"{a}nnstr\"{o}m, ``Precautionary
  safety for autonomous driving systems: Adapting driving policies to satisfy
  quantitative risk norms,'' in \emph{{IEEE} Intell. Transp. Syst. Conf.},
  2021, pp. 645--652.

\bibitem[Warg et~al.(2020)Warg, Skoglund, Thors\'{e}n, Johansson,
  Br\"{a}nnstr\"{o}m, Gyllenhammar, and Sanfridson]{WarSkoThoJohBraGyl:20}
F.~Warg, M.~Skoglund, A.~Thors\'{e}n, R.~Johansson, M.~Br\"{a}nnstr\"{o}m,
  M.~Gyllenhammar, and M.~Sanfridson, ``The quantitative risk norm - a proposed
  tailoring of {HARA} for {ADS},'' in \emph{50th Annual {IEEE/IFIP} Int. Conf.
  Depend. Syst. Netw. Workshops}, 2020, pp. 86--93.

\bibitem[Peden et~al.(2004)Peden, Scurfield, Sleet, Mohan, Hyder, Jarawan, and
  Mathers]{PedScuSleMohHydJar:04}
M.~Peden, R.~Scurfield, D.~Sleet, D.~Mohan, A.~A. Hyder, E.~Jarawan, and
  C.~Mathers, \emph{World report on road traffic injury prevention}.\hskip 1em
  plus 0.5em minus 0.4em\relax World Health Organization, Feb. 2004.

\bibitem[Kwiatkowska et~al.(2022)Kwiatkowska, Norman, and Parker]{KwiNorPar:22}
M.~Kwiatkowska, G.~Norman, and D.~Parker, ``Probabilistic model checking and
  autonomy,'' \emph{Annual Rev. Control, Robot. Auton. Syst.}, vol.~5, pp.
  385--410, 2022.

\end{thebibliography}

\end{document}